\documentclass[twocolumn,prl,showpacs,superscriptaddress]{revtex4}
\usepackage{graphicx}
\usepackage{dcolumn}
\usepackage{bm}

\begin{document}

\title{Probing non-Abelian statistics of Majorana fermions in ultracold atomic superfluid}

\author{Shi-Liang Zhu}
\affiliation{Laboratory of Quantum Information Technology and
SPTE,South China Normal University, Guangzhou, China}

\author{L.-B. Shao}
\affiliation{Department of Physics and Center of Theoretical and
Computational Physics, The University of Hong Kong, Pokfulam Road,
Hong Kong, China}

\author{Z. D. Wang}
\affiliation{Department of Physics and Center of Theoretical and
Computational Physics, The University of Hong Kong, Pokfulam Road,
Hong Kong, China}

\author{L.-M. Duan} \affiliation{Department of Physics and MCTP,
Michigan University, Ann Arbor, USA}

\begin{abstract}
We propose an experiment to directly probe the non-Abelian
statistics of Majorana fermions by braiding them in an s-wave
superfluid of ultracold atoms. We show different orders of
braiding operations give orthogonal output states that can be
distinguished through Raman spectroscopy. Realization of Majorana
states in an s-wave superfluid requires strong spin-orbital
coupling and a controllable Zeeman field in the perpendicular
direction. We present a simple laser configuration to generate the
artificial spin-orbital coupling and the required Zeeman field in
the dark state subspace.
\end{abstract}

\maketitle

Quantum statistics is a fundamental concept in physics that distinguishes
fermions from bosons. The discovery of the fractional quantum Hall (FQH)
effects opens up the possibility to study exotic quasi-particles beyond
bosons and fermions, which are called anyons \cite{Laughlin,Wilczek}.
There are two types of anyons, with Abelian or non-Abelian statistics. For
Abelian anyons, the winding of one quasi-particle around another gives a
phase factor to the wave function which is a fraction of $2\pi $. For
Non-Abelian anyons, a winding of one quasi-particle around another is
associated with a unitary transformation in the subspace of degenerate
ground states \cite{Moore}. Such unitary transformations form a
representation of the braid group where the elements in general are
non-commutative. Recently, the systems with non-Abelian anyons have
attracted considerable interest as candidates for realization of
fault-tolerant quantum computation \cite{kitaev,Nayak}.

Probing quantum statistics is a challenge for both Abelian and non-Abelian
anyons. Recent theoretical works suggest that interference experiments in a
well designed point contact geometry may serve as a probe of Abelian \cite%
{Chamon} or non-Abelian anyons \cite{na} in the FQH system.
Besides of the quasi-particles in the FQH system, a vortex
excitation in $p_{x}+ip_{y}$ superconductor traps a zero-energy
bound state which represents a kind of self-conjugate particles
called Majorana fermions \cite{Read}. Majorana fermions have the
feature of non-Abelian statistics and have raised
significant interest in condensed matter physics \cite%
{Nayak,Read,Tewari,Ivanov,Sau,Fu}. Several inspiring proposals have been
suggested very recently to realize Majorana fermions in different materials
\cite{Fu,Sau}. A scheme to observe Majorana fermions has also been
proposed in the p-wave superfluid state of ultracold atoms \cite{Tewari}.

In this paper, we propose an experiment to probe non-Abelian
statistics of the Majorana fermions with an ultracold atomic gas
in the s-wave superfluid state. In contract to the p-wave
superfluid state, which still remains to be realized
experimentally due to challenge associated with instability of the
p-wave Feshbach resonance \cite{Jin}, the s-wave superfluid state
is stable and has been observed in many experiments
\cite{Ketterle}. It has been noted recently that an s-wave
superfluid with spin-orbital (SO) coupling emulates properties of
the p-wave superfluid \cite{Fu,Zhang,Sau,Sato}. Realization of
Majorana fermions in an s-wave superfluid state requires SO
coupling and a controllable Zeeman field on the effective spins
\cite{Sau}. For ultracold atoms, an effective SO coupling can be
generated
through gradient of spatially varying Berry phase induced from laser beams \cite%
{SO1,SO2, Zhang2010}. We propose a scheme to realize both SO
coupling and the required perpendicular Zeeman field on the
effective spin, which a laser configuration compatible with the
current experiment and significantly simpler than the previous
proposal \cite{Zhang2010}. The Majorana fermions can be
adiabatically moved around each other through control of the
pinning potentials from focused laser beams. We construct
explicitly two composite braiding operations $A$ and $B$, for
which the different orders of braiding operations $AB$ and $BA$,
acting on the initial vacuum, give orthogonal output states that
can be distinguished through the Raman spectroscopy. The proposed
experiment thus provides a promising approach to directly probe
the fundamental non-Abelian statistics of the anyons. Realization
of Majorana fermions has been proposed in other systems
\cite{Fu,Sau}, however, for typical condensed matter materials, it
is very challenging to directly demonstrate the non-Abelian
statistics of the Majorana fermions by braiding them in space. The
proposed realization of Majorana fermions in an s-wave atomic
superfluid therefore offers a unique opportunity to adiabatically
rotate them in space with focused laser beams to observe the
non-Abelian fractional statistics, which is the most interesting
property associated with these exotic quasi-particles.

We consider a cloud of
ultracold fermionic atoms with three relevant spin components in
the ground-state manifold, label as $|1\rangle ,|2\rangle
,|3\rangle $ (e.g., they can be taken as the lowest three Zeeman
levels of $^{6}$Li atoms near the broad s-wave Feshbach
resonance).
To generate the effective SO
coupling, we apply three traveling wave laser beams in the $x$ and
$y$ directions which couple the atomic levels $|1\rangle
,|2\rangle ,|3\rangle $ resonantly to the excited state $|e\rangle
$ through the standard tripod configuration as shown in Fig. 1
\cite{SO1,SO2, Zhang2010}.
The single-particle Hamiltonian for each atom takes the form $H_{s}=\mathbf{p%
}^{2}/2m+H_{ej}$, where $\mathbf{p}$ denotes the momentum operator
and $m$ is the atomic mass. The light-atom interaction Hamiltonian
$H_{ej}$ is given by $H_{ej}=\hbar \sum_{j=1}^{j=3}(\Omega
_{j}|e\rangle \langle j|+h.c.)=\hbar \Omega \,|e\rangle
\left\langle B\right\vert +h.c.$, where the corresponding Rabi
frequencies take the form $\Omega _{1}=\Omega \,\sin
\beta \,\cos \phi \,\mathrm{e}^{-ikx},$ $\Omega _{2}=\Omega \,\cos \beta \,%
\mathrm{e}^{-iky},$ and $\Omega _{3}=\Omega \,\sin \beta \,\sin \phi \,%
\mathrm{e}^{ikx}$ with $k$ the wave number of the light, $\Omega
=\sqrt{|\Omega _{1}|^{2}+|\Omega _{2}|^{2}+|\Omega _{3}|^{2}}$,
and the bright state $|B\rangle \equiv \left( \Omega _{1}^{\ast
}|1\rangle +\Omega _{2}^{\ast }|2\rangle +\Omega _{3}^{\ast
}|3\rangle \right) /\Omega $. We assume uniform plane wave laser
beams so that $\Omega ,\beta ,\phi $ are all constants and the values of $%
\beta ,\phi $ will be specified below. The diagonalization of
$H_{ej}$
yields two orthogonal dark states $|D_{1}\rangle =\sin \phi \mathrm{e}%
^{-ik(y-x)}|1\rangle -\cos \phi \mathrm{e}^{-ik(y+x)}|3\rangle ,$ $%
|D_{2}\rangle =\cos \beta \cos \phi \mathrm{e}^{-ik(y-x)}|1\rangle
-\sin \beta |2\rangle +\cos \beta \sin \phi
\mathrm{e}^{-ik(y+x)}|3\rangle ,$ which are not coupled to the
excited state $|e\rangle $ and thus immune to spontaneous
emission. We assume the atoms are initially pumped to these dark
states and they remain in the dark states as long as the typical
velocity $v$ of the atoms satisfy the condition $\eta vk\ll \Omega
$ \cite{note}, where the prefactor $\eta \approx 1.3$ for the
isotropic  SO coupling discussed below. In typical experiments,
the wave number $k \approx 10^7\ m^{-1}$ and the Rabi frequency
$\Omega$ of the order of magnitude $2\pi \times 20 $ MHz can be
readily obtained, so the condition $\eta vk\ll \Omega $ is fully
satisfied for ultracold atoms with the typical atomic velocity
about several centimeters per second. The probability of
non-adiabatic excitation, estimated by $(\eta vk / \Omega)^2 $, is
negligible (about $10^{-6}$). In a resonant coupling configuration
as shown in Fig. 1a, the effective atomic spontaneous emission
rate is reduced from the original rate $\gamma_s$  by a factor of
the non-adiabatic excitation probability and thus estimated by
$\gamma_s(\eta vk / \Omega)^2 $, where is at a few Hz level. So
the atomic spontaneous emissions induced by these laser beams are
negligible if the experimental time is within $100$ ms. The
effective atomic spontaneous emission rate can be further reduced
if one uses off-resonant coupling configuration or increases the
Rabi frequency $\Omega$ with more laser power. Under the above
condition, we can project the Hamiltonian $H_{s}$ to the subspace
spanned by these two dark states, and the projected Hamiltonian
takes the form
\begin{equation}
H_{so}=\frac{(p_{x}-A_{x})^{2}}{2m}+\frac{(p_{y}-A_{y})^{2}}{2m},
\label{H_SO}
\end{equation}%
where $A_{x}=-\hbar k\cos \beta \sigma _{x}$, $A_{y}=\left( \hbar k/2\right)
\sin ^{2}\beta \sigma _{y}+\left( \hbar k/2\right) (1+\cos ^{2}\beta )$. In
the derivation, we have assumed $\phi =\pi /4$ and defined an effective spin
with the basis vector $|\uparrow \rangle =(|D_{1}\rangle +i|D_{2}\rangle )/%
\sqrt{2}$ and $|\downarrow \rangle =(i |D_{1}\rangle+|D_{2}\rangle
)/\sqrt{2} $. The Pauli matrices $\sigma _{x},\sigma _{y}$ are
defined in association with these spin basis vectors. Equation
(\ref{H_SO}) describes the atom in a spin-dependent synesthetic
gauge field \cite{SO1,SO2, Zhang2010}, and a spin-independent
version of this gauge field has been demonstrated in recent
experiments \cite{Lin}. To get the standard form of the SO
coupling, we choose the value of $\beta $ to satisfy $\sin
^{2}\beta =2\cos \beta =2(\sqrt{2}-1)$ and observe the system in
the moving frame with the velocity $v_{y}=\left( \hbar k/2m\right)
(1+\cos ^{2}\beta )$,
so that we can drop the \ constant term $\left( \hbar
k/2\right) (1+\cos ^{2}\beta )$ in $A_{y}$.

The effective Hamiltonian $H_{so}$ has the required SO coupling,
but we still need to generate a perpendicular Zeeman field on the
effective
spin $|\uparrow \rangle ,|\downarrow \rangle $ with the coupling term $%
h_{z}\sigma _{z}$. For this purpose, we apply two additional laser beams
that couple the states $|1\rangle $ and $|3\rangle $ to the excited state $%
|e\rangle $ off-resonantly with a large detuning $\Delta_d $ (see
Fig.1(a)), with the corresponding Rabi frequencies denoted by
$\Omega _{1z}=i\left\vert \Omega _{z}\right\vert
\mathrm{e}^{-ikx}$ and $\Omega _{3z}=\left\vert \Omega
_{z}\right\vert \mathrm{e}^{ikx}$. With the driving $\left\vert
\Omega _{1z}\right\vert ,\left\vert \Omega _{3z}\right\vert \ll
\Delta_d $, it leads to a perturbation Hamiltonian given by
$H_{z}=-i\hbar \Omega _{p}e^{2ikx}|1\rangle \langle 3|+h.c$, where
$\Omega _{p}=|\Omega _{z}|^{2}/\Delta_d $ \cite{note2}. We assume
$\Omega _{p}\ll \Omega $, so the Hamiltonian $H_{p}$ can not pump
the atoms outside of the dark state space spanned by the
$|\uparrow \rangle $ and $|\downarrow \rangle $ levels, however,
it splits the degeneracy of these two levels and leads to a
coupling Hamiltonian of the form $H_{z}=\left( \hbar \Omega
_{p}\cos \beta \right) \sigma _{z}$ in the dark state space, which
gives exactly the effective Zeeman field term $h_{z}\sigma _{z}$
with $h_{z}=\hbar \Omega _{p}\cos \beta $. The Zeeman field
$h_z/\hbar$ can be continuously tuned from zero to about $2\pi
\times 50 $ kHz (enough for our later applications), and under
this condition, the requirement $\Omega _{p}\ll \Omega $ is well
satisfied with the off-resonant excitation probability $(\Omega
_{p}/ \Omega)^2 < 10^{-5}$. Note that in this scheme we need only
two laser beams shined from the same direction as $\Omega
_{1},\Omega _{2}$ to create the required Zeeman field. This
significantly simplifies the experimental requirement compared
with the scheme in Ref. \cite{Zhang2010} which needs five
additional laser beams and more complicated optical configuration
to create the Zeeman field.

\begin{figure}[tbph]
\label{Fig1} \includegraphics[height=3.5cm]{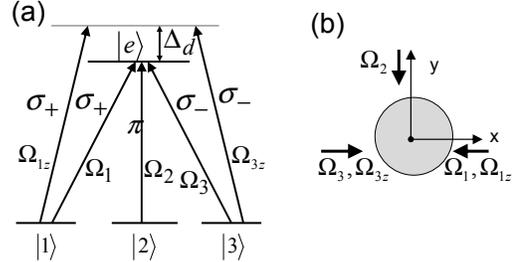}
\caption{Schematic representation of the light-atom interaction
configuration for generation of the effective Hamiltonian with SO
coupling and a perpendicular Zeeman field. (a) The relevant atomic
levels and the coupling Rabi frequencies to corresponding laser
beams with specified polarizations $\sigma_+$, $\sigma_-$, or
$\pi$. (b) The spatial configuration and propagating direction of
the laser beams.}
\end{figure}

Now we have the effective single-particle Hamiltonian $H_{so}+H_{z}$ of the
right form. The atomic interaction is described by the s-wave scattering. In
the second quantization formalism, it is described by the Hamiltonian $%
H_{int}=\sum_{i<j}\int g_{ij}\Psi _{i}^{\dag }\left(
\mathbf{r}\right) \Psi _{j}^{\dag }\left( \mathbf{r}\right) \Psi
_{j}\left( \mathbf{r}\right) \Psi _{i}\left( \mathbf{r}\right)
\mathbf{d}^{3}\mathbf{r}$, where $\Psi _{i}\left(
\mathbf{r}\right) $ denote the fermionic field operators, $i,j$
are summarized over the three spin components $1,2,3$, and
$g_{ij}$ are proportional to the s-wave scattering lengths in the
corresponding channel. In the s-wave superfluid state, the
Hamiltonian $H_{int}$ can be replaced by the effective mean-field
BCS type Hamiltonian $H_{BCS}=\sum_{i<j}\int \Delta _{ij}\left(
\mathbf{r}\right) \Psi _{j}\left( \mathbf{r}\right) \Psi
_{i}\left( \mathbf{r}\right) \mathbf{d}^{3}\mathbf{r+h.c.}$, where
$\Delta _{ij}\left( \mathbf{r}\right) \equiv g_{ij}\left\langle
\Psi _{i}^{\dag }\left( \mathbf{r}\right) \Psi _{j}^{\dag }\left(
\mathbf{r}\right) \right\rangle $ correspond to the superfluid
order parameters. The typical s-wave pairing potential
$\Delta_{ij}$ in experiments is about $2\pi \times 3.5 $ kHz
\cite{Ketterle}, which is much smaller than  $\Omega $,
 so the interaction Hamiltonian $H_{BCS}$ can not pump the atoms
outside of the dark-state subspace. Under this condition, we can
project $H_{BCS}$ to the dark-state subspace, and the projected
Hamiltonian $H_{P}$ takes the form $H_{P}=\int \Delta \left( \mathbf{r}%
\right) \Psi _{\uparrow }\left( \mathbf{r}\right) \Psi _{\downarrow }\left(
\mathbf{r}\right) \mathbf{d}^{3}\mathbf{r+h.c.}$, where $\Delta $ is a
linear superposition of $\Delta _{12},\Delta _{13},\Delta _{23}$ with the
superposition coefficients determined from the transition matrix between $%
|1\rangle ,|2\rangle ,|3\rangle $ and $|\uparrow\rangle$,
$|\downarrow\rangle$, $|B\rangle$. If we write the wave-function $\Phi (\mathbf{r})$ as $%
\Phi (\mathbf{r})=[u_{\uparrow }(\mathbf{r}),u_{\downarrow }(\mathbf{r}%
),v_{\downarrow }(\mathbf{r}),v_{\uparrow }(\mathbf{r})]^{T}$ in
the basis specified by the operators $(\Psi _{\uparrow
}(\mathbf{r}),\Psi _{\downarrow }(\mathbf{r}),\Psi _{\downarrow
}^{\dagger }(\mathbf{r}),-\Psi _{\uparrow }^{\dagger
}(\mathbf{r}))^{T}$, the Schrodinger equation for the total
Hamiltonian $H=H_{so}+H_{z}+H_{P}$ with eigen-energy $E$ takes the
form of the Bogoliubov-de Gennes (BdG) equation%
\begin{equation}
\left(
\begin{array}{cc}
H_{so}+H_{z} & \Delta (\mathbf{r}) \\
\Delta ^{\ast }(\mathbf{r}) & -\sigma _{y}\left( H_{so}^{\ast }+H_{z}^{\ast
}\right) \sigma _{y}%
\end{array}%
\right) \Phi (\mathbf{r})=E\Phi (\mathbf{r}).
\end{equation}%
If we introduce $\tau _{x,y,z}$ to denote Pauli matrices in the Nambu spinor
space, the Hamiltonian $H$ can be written in the form

\begin{eqnarray}
\label{H_single} H &=&[\frac{p^{2}}{2m}-\mu ]\tau _{z}+\alpha
(\sigma _{x}p_{x}+\sigma
_{y}p_{y})\tau _{z}+h_{z}\sigma _{z}  \nonumber \\
&&+(\tau _{x}\text{Re}\Delta -\tau _{y}\text{Im}\Delta ),  \label{Ht}
\end{eqnarray}%
where the laser induced SO coupling rate $\alpha =-\hbar k(\sqrt{2}%
-1).$ All the constant potentials have been included into the definition of
the chemical potential $\mu $ which is self-consistently determined from the
particle density. For ultracold atoms in a global spin-independent confining
trap $V(\mathbf{r})$ , under the local density approximation, the global
trap can also be included into the definition of the chemical potential with
the substitution $\mu \rightarrow \mu -V(\mathbf{r})$.

The Hamiltonian $H$ in Eq.(\ref{H_single}) has been investigated
in the context of a recent proposal for realization of non-Abelian
anyons in a solid state heterostructure consisting of a
superconductor, a 2D electron gas with Rashba SO coupling, and a
magnetic insulator \cite{Sau}. It is known that for this
Hamiltonian, a Majorana fermion zero mode emerges
from a vortex core if the parameters in the Hamiltonian satisfy the condition $%
h_{z}>h_c=\sqrt{\mu^{2}+\left\vert \Delta \right\vert ^{2}}$
\cite{Sau,Sato}. For ultracold atoms, $h_c $ is comparable with
the Fermi energy $E_f$ since the chemical potential $|\mu| \sim
E_f$ and the pairing potential $|\Delta| \sim 0.35 E_f$. The
condition $h_z > h_c$ can be easily satisfied since we assume
$h_z/\hbar$ can be tuned up to $2\pi\times 50$ kHz, while
$E_f/\hbar$ is around $2\pi \times 10$ kHz \cite{Ketterle}.  A
vortex is needed for emergency of the Majorana fermion in a
central potential because the Hamiltonian has the particle-hole
symmetry \cite{Sau}. Vortices in atomic superfluid can be created
with several methods and have been observed in many experiments
\cite{vortex1,vortex2}. For instance, the required angular
momentum associated with the vortices can come from initial
rotation of the atomic cloud \cite{vortex1} or be imprinted from
nonzero orbital angular momentum of laser beams \cite{vortex2}.

We now turn to the important question of probing non-Abelian
statistics associated with Majorana fermions.
For an unambiguous demonstration of the
non-Abelian statistics, we need to construct two braiding
operations $A$ and $B$. When applied to the initial state, the
composite braidings $AB$ and $BA$ give orthogonal output states
which can be distinguished through the experimental detection. In
the following, we explicitly construct these braiding operations
for Majorana fermions and discuss the experimental detection of
the output states.

Let us consider $2N$ Majorana fermions confined in $2N$ vortex
cores, denote by $\gamma _{i}$ $(i=1,2,\cdots ,2N)$. The Majorana
fermions can be combined pairwise to form $N$ complex fermionic
states $c_{j}=\gamma _{2j}+i\gamma _{2j-1}$. Each fermionic state
can be occupied or unoccupied and therefore the ground state is
$2^{N}$-fold degenerate. If we exchange one Majorana fermion $j$
with its nearest neighbor $j+1$ by adiabatically moving the
pinning potential, we get an unitary operation $\tau _{j}=\exp
\left( \pi \gamma _{j+1}\gamma _{j}/4\right) =(1+\gamma
_{j+1}\gamma _{j})/\sqrt{2}$ acting on the $2^{N}$-dimensional
degenerate Hilbert space \cite{Ivanov}. The elementary interchange
operations $\tau _{j}$ $(j=1,2,\cdots ,2N-1)$\ form a complete set
of generators for the braid group. To demonstrate the non-Abelian
statistics, we need at least four Majorana fermions due to
existence of the superselection rule. For four Majorana fermions,
they combine into two complex fermions $c_{1},c_{2}$. The ground
state has degeneracy four, but the Hamiltonian $H$ conserves
parity of the fermion number, and the even-number subspace spanned
by $\{|0\rangle ,c_{1}^{\dagger }c_{2}^{\dagger }|0\rangle \}$ is
decoupled from the odd-number subspace spanned by
$\{c_{1}^{\dagger }|0\rangle ,c_{2}^{\dagger }|0\rangle \}$. At
low temperature (temperature significantly below the superfluid
gap), the initial state is typically a vacuum state $|0\rangle $
\cite{Tewari}, and in the even-number subspace $\{|0\rangle
,c_{1}^{\dagger }c_{2}^{\dagger }|0\rangle \}$, only two of the
three generators $\tau _{1},\tau _{2},\tau _{3}$ are independent
with
\begin{equation}
\tau _{1}=\tau _{3}=\frac{1}{\sqrt{2}}\left(
\begin{array}{cc}
1-i & 0 \\
0 & 1+i%
\end{array}%
\right) ,\ \tau _{2}=\frac{1}{\sqrt{2}}\left(
\begin{array}{cc}
1 & -i \\
-i & 1%
\end{array}%
\right) .  \label{NonAbelian}
\end{equation}%
From these generators, we construct two composite braiding operations
\begin{equation}
A=\tau _{1}\tau _{2}\tau _{1}^{-1},\ B=\tau _{1}^{-1}\tau _{2}\tau _{1}^{-1}
\label{AB}
\end{equation}%
with the property
\begin{equation}
AB=i\sigma _{z},\ \ \ BA=-i\sigma _{x},
\end{equation}%
where $\tau _{1}^{-1}$ represents a braiding along the reverse
direction. With the initial state $|0\rangle $, the braiding
operations $AB$ and $BA$ yield two orthogonal output states,
either $i|0\rangle $ or $-ic_{1}^{\dagger }c_{2}^{\dagger
}|0\rangle $, demonstrating the non-Abelian nature of the anyons.
During braiding, we need to fulfill the adiabaticity. The gap to
the first excited state in the vortex core is estimated by
$\omega_0 \sim \Delta^2/E_f \sim 2\pi\times 1.2$ kHz. If the
braiding time $t_b$ is of the order of tens of ms, the
non-adiabatic excitation probability estimated by $1/(\omega_0
t_b)^2$ is negligible.

The two output states $i|0\rangle $ and $-ic_{1}^{\dagger
}c_{2}^{\dagger }|0\rangle $ can be distinguished through the
Raman (or r.f.) spectroscopy \cite{Schirotzek}. For the state
$c_{1}^{\dagger }c_{2}^{\dagger }|0\rangle $, we have two
fermionic atoms at zero energy trapped at the vortex cores. We can
apply a Raman $\pi $-pulse to resonantly bring these zero-energy
atoms to a different hyperfine state $|D\rangle $ \cite{Tewari},
which is outside of Feshbach resonance and thus only weakly
interacting with other atoms.
If we choose the Rabi frequency of the Raman pulse to have a
Gaussian shape $\Omega_R=\Omega_0 \exp(-r^2/w^2) \exp(-\omega_0^2
t^2)$ (where $-t_f \le t \le t_f$ and the center of the vortex
core is at $r=0$), it will not break up the Cooper-pairs or excite
other quasiparticles at the
vortex core which are detuned typically by an energy scale comparable with $%
\left\vert \Delta \right\vert $. It is estimated that the error is
about $4 \times 10^{-4}$ and thus negligible under typical
experimental parameters $\Delta=2\pi \times 3.5$ kHz,
$\omega_0=0.35\Delta$, $\Omega_0=1.7\omega_0$, $t_f=5/\omega_0$
and $w=1.5 \mu m$. If the two laser beams for the Raman pulse
propagate along different directions, it also applies a momentum
kick to the target atoms to push them outside of the atomic cloud.
These target atoms in a different hyperfine state $|D\rangle $ can
then be detected with the standard quantum jump technique.

In summary, we have proposed an experimental scheme to create and
observe Majorana fermions in an s-wave superfluid of ultracold
fermionic atoms with laser induced SO coupling. The non-Abelian
statistics of the Majorana fermions can be directly probed through
laser-controlled braiding and detection based on the Raman
spectroscopy.

This work was supported by the NSF of China (No 10974059), the
SKPBR of China (Nos.2011CB922104 and 2007CB925204), the RGC of
Hong Kong, the DARPA OLE Program under ARO Award W911NF0710576,
and the AFOSR MURI program.


\begin{thebibliography}{99}
\bibitem{note} This condition is obtained by requiring the off-diagonal
coupling matrix element of the Hamiltonian $H_{s}$ between the
dark states and the bright state much smaller than the energy
difference $\hbar \Omega $.
\bibitem{note2} There is also a constant a.c.-Stark shift term $\left(
|\Omega _{z}|^{2}/\Delta_d \right) \left( |1\rangle \left\langle
1\right\vert +|3\rangle \left\langle 3\right\vert \right) $
associated with the laser beams $\Omega _{1z},\Omega _{3z}$.
However, this constant frequency shift can be easily canceled with
a corresponding frequency offset of the laser beam $\Omega _{2}$
applied to the level $|2\rangle $, so it has no physical effect on
the ground levels.
\end{thebibliography}
\end{document}